\documentclass[12pt]{article}

\newcommand{\beq}{\begin{equation}}
\newcommand{\eeq}{\end{equation}}
\newcommand{\beqa}{\begin{eqnarray}}
\newcommand{\eeqa}{\end{eqnarray}}
\newcommand{\beqar}{\begin{eqnarray*}}
\newcommand{\eeqar}{\end{eqnarray*}}

\newcommand{\eg}{{\it e.g.,}\ }
\newcommand{\ie}{{\it i.e.,}\ }
\newcommand{\labell}[1]{\label{#1}} 
\newcommand{\reef}[1]{(\ref{#1})}
\newcommand\prt{\partial}

\newcommand\cR{{\cal R}}
\newcommand\cL{{\cal L}}

\newcommand\hR{\hat{R}}

\newcommand\tG{{\widetilde G}}
\newcommand\tg{{\widetilde g}}

\newcommand\ta{{\tilde a}}
\newcommand\tb{{\tilde b}}
\newcommand\tc{{\tilde c}}
\newcommand\td{{\tilde d}}
\newcommand\te{{\tilde e}}

\newcommand\tf{{\tilde f}}

\begin{document}

\begin{titlepage}

\begin{center}



\vskip 2 cm
{\LARGE \bf T-duality of O-plane action\\  at order $ \alpha'^2$
 }\\
\vskip 1.25 cm
 Mohammad R. Garousi\footnote{garousi@ um.ac.ir}  \\
 \vskip 1cm
\vskip 1 cm
{{\it Department of Physics, Ferdowsi University of Mashhad\\}{\it P.O. Box 1436, Mashhad, Iran}\\}
\vskip .1 cm
\vskip .1 cm

\end{center}

\vskip 0.5 cm

\begin{abstract}
\baselineskip=18pt
We use compatibility of  O-plane  action  with nonlinear T-duality as a guiding principle to find all NS-NS couplings in the O-plane action at order  $\alpha'^2$. We find that the dilaton couplings appear in the string frame  only via the transformation $\hat{R}_{\mu\nu}\rightarrow \hat{R}_{\mu\nu}+\nabla_{\mu}\nabla_{\nu}\phi$. 
\end{abstract}

\end{titlepage}
\section{Introduction and Result}
The  low energy effective field theory of  D-branes  in type II superstring theories consists of the  Dirac-Born-Infeld (DBI) \cite{Bachas:1995kx} and the  Chern-Simons (CS) actions \cite{Douglas:1995bn}. The effective theory of  O-plane are  orientifold projection of above actions. The curvature corrections to the CS part  have been found in  \cite{Green:1996dd,Cheung:1997az,Minasian:1997mm} by requiring that the chiral anomaly on the world volume of intersecting D-branes (I-brane) cancels with the anomalous variation of the CS action. 
 The curvature corrections to the DBI  action, on the other hand,  have been found in \cite{Bachas:1999um} by requiring consistency of the effective action with the $O(\alpha'^2)$ terms of the corresponding disk-level scattering amplitude \cite{Garousi:1996ad,Hashimoto:1996kf}. For totally-geodesic embeddings of world-volume in the ambient spacetime, the corrections in the string frame  for zero B-field and for constant dilaton  are\footnote{Our index conversion is that the Greek letters  $(\mu,\nu,\cdots)$ are  the indices of the space-time coordinates, the Latin letters $(a,d,c,\cdots)$ are the world-volume indices and the letters $(i,j,k,\cdots)$ are the normal bundle indices.} \cite{Bachas:1999um}
\beqa
S &\!\!\!\!\supset\!\!\!\!&\frac{\pi^2\alpha'^2T_{p}}{48}\int d^{p+1}x\,e^{-\phi}\sqrt{-\tG}\bigg[R_{abcd}R^{abcd}-2\hR_{ab}\hR^{ab}-R_{abij}R^{abij}+2\hR_{ij}\hR^{ij}\bigg]\labell{DBI}
\eeqa
where $\hR_{ab}= R^c{}_{acb}$, $\hR_{ij}= R^c{}_{icj}$ and $\tG=\det(\tG_{ab})$ where  $\tG_{ab}$  is the pull-back of the bulk metric onto the word-volume, \ie
\beqa
\tG_{ab}=\frac{\prt X^{\mu}}{\prt\sigma^a}\frac{\prt X^{\nu}}{\prt\sigma^b}G_{\mu\nu}
\eeqa
  The orientifold projection, projects out the Riemann curvature with odd number of transverse indices, so the above couplings are  the curvature couplings on the world volume of  both D-brane and O-plane  depending on the tension $T_p$ which is different for D-brane and O-plane.  

In the presence of non-constant dilaton, the couplings \reef{DBI} are not consistent with T-duality. For zero B-field, the compatibility with linear T-duality requires the following extension \cite{Garousi:2009dj,Garousi:2011fc}:
\beqa
S &\!\!\!\!\supset\!\!\!\!&\frac{\pi^2\alpha'^2T_{p}}{48}\int d^{p+1}x\,e^{-\phi}\sqrt{-\tG}\bigg[R_{abcd}R^{abcd}-2\cR_{ab}\cR^{ab}-R_{abij}R^{abij}+2\cR_{ij}\cR^{ij}\bigg]\labell{DBI2}
\eeqa
where $\cR_{\mu\nu}=\hR_{\mu\nu}+\nabla_{\mu}\nabla_{\nu}\phi$. The orientifold projection projects out dilaton with odd number of transverse derivatives, so the above couplings are  valid for both D-brane and O-plane actions.

In the presence of non-zero B-field, the  couplings   \reef{DBI2} are not consistent with the T-duality. Using the compatibility of  these couplings  with linear T-duality as a guiding principle, the quadratic B-field couplings at order $O(\alpha'^2)$ have been found  in \cite{Garousi:2009dj} to be 
\beqa
S &\!\!\!\!\supset\!\!\!\!&\frac{\pi^2\alpha'^2T_{p}}{48}\int d^{p+1}xe^{-\phi}\sqrt{-\tG}\bigg[\frac{1}{2}\nabla_aH_{bci}\nabla^aH^{bci}-\frac{1}{6}\nabla_aH_{ijk}\nabla^aH^{ijk}-\frac{1}{3}\nabla_iH_{abc}\nabla^iH^{abc}
\bigg]\labell{LDBI}
\eeqa
 The above couplings have been confirmed with the disk level S-matrix calculations \cite{Garousi:2009dj}. The orientifold projection projects out covariant derivatives of B-field  with even number of transverse indices, so again the above couplings are  valid for both D-brane and O-plane.

The quadratic couplings \reef{DBI2} and \reef{LDBI} are not however consistent with the full nonlinear T-duality transformations. In this paper, we are going to use the compatibility of the couplings \reef{DBI} with nonlinear T-duality as a guiding principle to find the couplings of O-plane to all massless NS-NS fields at order $\alpha'^2$. We find the quadratic  couplings \reef{DBI2} and \reef{LDBI} as well as the following higher order couplings:
\beqa
S &\!\!\!\!\supset\!\!\!\!&\frac{\pi^2\alpha'^2T_{p}}{48}\int d^{p+1}xe^{-\phi}\sqrt{-\tG}\bigg[H^{abi} H_{a}{}^c{}_i \cR_{bc} -\frac{3}{2} H^{abi} H_{ab}{}^j \cR_{ i j}+\frac{1}{2} H^{ijk} H_{ij}{}^l\cR_{ k l}\nonumber\\&&-H^{abi} H^{cd}{}_i R_{abcd}+H^{abi} H_i{}^{jk} R_{abjk}-\frac{1}{4} H^{abi} H_{ab}{}^j H_i{}^{kl} H_{jkl}+\frac{1}{4} H^{abi} H_{ab}{}^j H^{cd}{}_i H_{cdj}\nonumber\\&& +\frac{1}{8} H^{abi} H_{a}{}^{cj} H_b{}^d{}_j H_{cdi}-\frac{1}{6} H^{abi} H_{a}{}^{cj} H_{bc}{}^k H_{ijk}+\frac{1}{24} H^{ijk} H_i{}^{lm} H_{jl}{}^n H_{kmn}\bigg]\labell{finalH}
\eeqa
 The consistency of the effective actions with   T-duality has been also  used in \cite{Myers:1999ps}-\cite{Garousi:2013gea} to find new couplings in the world volume and spacetime actions. 

An outline of the paper is as follows: In the next section, we present an algorithm for calculating the world volume theory of D-brane/O-plane by imposing the action to be consistent with the T-duality transformations. In section 3, we find the couplings of gravity and dilaton for O-plane which are invariant under a simplified T-duality transformations in which there is no B-field and the metric is diagonal. We find six  multiplets which are invariant under the simplified T-duality transformations. These multiplets, however, are not  invariant under the full T-duality transformations. Using the consistency of the multiplets with S-matrix elements, we argue that only two  of these multiplets survive under the full T-duality transformations. In section 4, we find the appropriate B-field couplings which make the two multiplets to be invariant under the full T-duality transformations. 

\section{T-duality Constraint }

The full set of nonlinear T-duality transformations for massless  fields have been found in \cite{TB,Meessen:1998qm,Bergshoeff:1995as,Bergshoeff:1996ui,Hassan:1999bv}.   When  the T-duality transformation acts along the Killing coordinate $y$,  the transformations of NS-NS fields are
\beqa
&&\,\,e^{2\phi}\rightarrow\frac{e^{2\phi}}{G_{yy}}\,,\quad\quad\,
G_{yy}\rightarrow\frac{1}{G_{yy}}\,,\nonumber\\
&&G_{\alpha y}\rightarrow\frac{B_{\alpha y}}{G_{yy}}\,,\qquad
G_{\alpha\beta}\rightarrow G_{\alpha\beta}-\frac{G_{\alpha y}G_{\beta y}-B_{\alpha y}B_{\beta y}}{G_{yy}}\,,\nonumber\\
&&B_{\alpha y}\rightarrow\frac{G_{\alpha y}}{G_{yy}}\,,\qquad
B_{\alpha\beta}\rightarrow B_{\alpha\beta}-\frac{B_{\alpha y}G_{\beta y}-G_{\alpha y}B_{\beta y}}{G_{yy}} \,,\labell{Cy1}
\eeqa
where $\alpha,\beta\ne y$. In above transformation the metric is given in the string frame. If $y$ is identified on a circle of radius $\rho$, \ie $y\sim y+2\pi \rho$, then after T-duality the radius becomes $\tilde{\rho}=\alpha'/\rho$. The string coupling is also transformed as $\tilde{g}=g\sqrt{\alpha'}/\rho$. It is known that the above transformations do not receive $\alpha'$ correction in the type II superstring theories in which we are interested.

If one defines the new field $\varphi$ as $G_{yy}=e^{-\varphi}$ and uses the dimensional reduction to write the 10-dimensional metric and B-field as 
\beqa
G_{\mu\nu}=\left(\matrix{g_{\alpha\beta}+e^{\varphi}g_{\alpha }g_{\beta }& e^{\varphi}g_{\alpha }&\cr e^{\varphi}g_{\beta }&e^{\varphi}&}\right)\,,\qquad B_{\mu\nu}= \left(\matrix{b_{\alpha\beta}+\frac{1}{2}b_{\alpha }g_{\beta }- \frac{1}{2}b_{\beta }g_{\alpha }&b_{\alpha }\cr - b_{\beta }&0&}\right)\labell{reduc}\eeqa
where $g_{\alpha\beta}, \,b_{\alpha\beta}$ are the metric and the B-field, and $g_{\alpha},\, b_{\alpha}$ are two vectors  in the 9-dimensional base space, then the T-duality transformations \reef{Cy1} simplify to 
\beqa
\phi\rightarrow\phi-\frac{1}{2}\varphi\,\,\,,\,\,\varphi\rightarrow -\varphi
\,\,\,,\,\,g_{\alpha }\rightarrow b_{\alpha }\,\,\,,\,\, b_{\alpha }\rightarrow g_{\alpha } \labell{T2}
\eeqa
The 9-dimensional base space fields $g_{\alpha\beta}$ and $b_{\alpha\beta}$ remain invariant under the T-duality.

A method for finding the world volume couplings which are invariant under linear T-duality is given in \cite{Garousi:2009dj}. This method may be used   to show that the S-matrix elements satisfy the  Ward identity corresponding to the T-duality (see \eg  \cite{Velni:2013jha}). In this section, we are going to extend this  method to find the  world volume action at   order   $\alpha'^2$ which is invariant under nonlinear T-duality. To this end, we first write all covariant couplings at   order $\alpha'^2$  with unknown coefficients. These couplings can be constructed by contracting the appropriate bulk tensors with the inverse of bulk metric $G^{\mu\nu}$ or with the world volume first fundamental form  $\tG^{\mu\nu}$. This tensors is defined as
\beqa
\tG^{\mu\nu}=\frac{\prt X^{\mu}}{\prt\sigma^a}\frac{\prt X^{\nu}}{\prt\sigma^b}\tG^{ab}
\eeqa
where $\tG^{ab}$ is inverse of the pull-back  metric  $\tG_{ab}$.
 We call the  action corresponding to these couplings  $S$. Then we reduce the action to the 9-dimensional space which depends on whether the Killing coordinate $y$ is a world volume or transverse direction.     When $y$ is a world volume direction   we  call the reduced action  $S^w$, and  when   $y$ is a transverse direction   we  call it $S^t$.  The T-duality of $S^w$ which we call it $S^{wT}$ must be equal to $S^t$ up to some total derivative terms, \ie
\beqa
  S^{wT}-S^t=0\labell{SS}
\eeqa
  The above constraint can be used to find the unknown coefficients in the original action $S$.

Using the reductions \reef{reduc} and the reduction of $G^{\mu\nu}$ which is the  inverse of $G_{\mu\nu}$ in  \reef{reduc}, \ie
\beqa
G^{\mu\nu}=\left(\matrix{g^{\alpha\beta} & - g^{\alpha }&\cr - g^{\beta  }&e^{-\varphi}+g_{\chi }g^{\chi}&}\right)\labell{reduc2}
\eeqa
where $g^{\alpha\beta}$ is inverse of $g_{\alpha\beta}$, it is straightforward to reduce the bulk tensors  which contract only with $G^{\mu\nu}$. One should first separate the 10-dimensional indices to the 9-dimensional indices and the $y$ index. Then one should reduce the 10-dimensional field in them  to the 9-dimensional fields according to \reef{reduc} and \reef{reduc2}.  

For the couplings in which the bulk tensors contract with $G^{\mu\nu}$ and  $\tG^{\mu\nu}$, we also need the reduction of  $\tG^{\mu\nu}$. The reduction of the first fundamental form depends on weather the brane is along or orthogonal to the circle. In the former case, the reduction of $\tG_{ab}$ is 
\beqa
\tG_{ab}=\left(\matrix{\frac{\prt X^{\alpha}}{\prt\sigma^\ta}\frac{\prt X^{\beta}}{\prt\sigma^\tb}g_{\alpha\beta}+e^{\varphi}\frac{\prt X^{\alpha}}{\prt\sigma^\ta}\frac{\prt X^{\beta}}{\prt\sigma^\tb}g_{\alpha }g_{\beta }& e^{\varphi}\frac{\prt X^{\alpha}}{\prt\sigma^\ta}g_{\alpha }&\cr e^{\varphi}\frac{\prt X^{\beta}}{\prt\sigma^\tb}g_{\beta }&e^{\varphi}&}\right) \labell{reducab}\eeqa
where the world volume indices $\ta\,,\tb\ne y$. Inverse of this matrix is
\beqa
\tG^{ab}=\left(\matrix{ \tg^{\ta\tb}& -  \tg^\ta&\cr -  \tg^\tb&e^{-\varphi}+\tg_{\ta }\tg^{\ta}&}\right) \labell{reducabI}\eeqa
where $\tg^{\ta\tb}$ is inverse of the pull-back of the 9-dimensional bulk metric onto the word-volume, \ie
\beqa
\tg_{\ta\tb}=\frac{\prt X^{\alpha}}{\prt\sigma^\ta}\frac{\prt X^{\beta}}{\prt\sigma^\tb}g_{\alpha\beta}
\eeqa
 and $\tg^\ta=\tg^{\ta\tb}\frac{\prt X^{\alpha}}{\prt\sigma^\tb}g_{\alpha }$. Therefore, the reduction of  $\tG^{\mu\nu}$ becomes
\beqa
\tG^{\mu\nu}=\left(\matrix{\frac{\prt X^{\alpha}}{\prt\sigma^\ta}\frac{\prt X^{\beta}}{\prt\sigma^\tb}\tg^{\ta\tb}& - \frac{\prt X^{\alpha}}{\prt\sigma^\ta}\tg^\ta&\cr -  \frac{\prt X^{\beta}}{\prt\sigma^\tb}\tg^\tb&e^{-\varphi}+\tg_{\ta }\tg^{\ta}&}\right)
\eeqa

In the latter case where $y$ is not a world volume index,  the reduction of $\tG_{ab}$ is 
\beqa
\tG_{ab}=\left(\matrix{\frac{\prt X^{\alpha}}{\prt\sigma^\ta}\frac{\prt X^{\beta}}{\prt\sigma^\tb}g_{\alpha\beta} & 0&\cr 0&0&}\right) \labell{reducabi}\eeqa
and its  inverse   is
\beqa
\tG^{ab}=\left(\matrix{ \tg^{\ta\tb}& 0&\cr 0&0&}\right) \labell{reducabIi}\eeqa
  Therefore, the reduction of  $\tG^{\mu\nu}$ in this case becomes
\beqa
\tG^{\mu\nu}=\left(\matrix{\frac{\prt X^{\alpha}}{\prt\sigma^\ta}\frac{\prt X^{\beta}}{\prt\sigma^\tb}\tg^{\ta\tb}& 0&\cr 0&0&}\right)
\eeqa

To proceed further, we need to   fix the world volume reparametrization invariance of the action. We fix it by choosing the static gauge where $X^a=\sigma^a$ and $X^i=2\pi\alpha'\Phi^i$. For O-plane at $X^i=0$ and D-brane at fixed position $X^i=0$, one finds $\frac{\prt X^{\alpha}}{\prt\sigma^\ta}=\delta^{\alpha}_{\ta}$. As a result,  the pull-back metric becomes $\tg_{\ta\tb}=g_{\ta\tb}$ and  the above reductions become
\beqa
\tG^{\mu\nu}=\left(\matrix{ g^{\ta\tb}& -  g^\ta&\cr - g^\tb&e^{-\varphi}+g_{\ta }g^{\ta}&}\right)\labell{tG1}
\eeqa
when brane is along the $y$ direction, and 
\beqa
\tG^{\mu\nu}=\left(\matrix{ g^{\ta\tb}& 0&\cr 0&0&}\right)\labell{tG2}
\eeqa
when brane is orthogonal to  the $y$ direction. Using the   reductions \reef{reduc} and \reef{reduc2}, and the reductions \reef{tG1}, \reef{tG2}, it is then straightforward    to reduce the bulk tensors which contract  with $G^{\mu\nu}$ and with the first fundamental form $\tG^{\mu\nu}$ .

The T-duality requires the world volume action to have the following structure:
\beqa
S=\int d^{p+1}xe^{-\phi}\sqrt{-\tG}\cL
\eeqa
Only the covariant derivatives of dilaton appears in  $\cL$.  The reduction of $e^{-\phi}\sqrt{- \tG}$ along a world volume direction  can be read from \reef{reducab} to be   $e^{-\phi+\varphi/2}\sqrt{- \tg}$ where $\tg=\det(\tg_{\ta\tb})$. It transforms to  $e^{-\phi}\sqrt{-\tg}$ under the T-duality \reef{T2}. On the other hand,  the reduction of  $e^{-\phi}\sqrt{-\tG}$ along a transverse direction can be read from \reef{reducabi} to be $e^{-\phi}\sqrt{-\tg}$. So the constraint \reef{SS} can be written as 
\beqa
\int d^p e^{-\phi}\sqrt{-\tg}\bigg[\cL^{wT}-\cL^t\bigg]=0\labell{LL}
\eeqa
where $\cL^{wT}$ is the T-duality of reduction of Lagrangian $\cL$ when brane is along the $y$-direction, and $\cL^t$ is the reduction of $\cL$ when brane is orthogonal to the $y$-direction.

To satisfy the constraint \reef{LL} there are two possibilities. One  is to consider all couplings with  arbitrary covariant derivatives in $\cL$ which are at order $\alpha'^2$, \eg $\nabla^2R$, and then to impose the constraint that the Lagrangian is T-duality invariant, \ie $\cL^{wT}-\cL^t=0$.  Another  possibility is to consider only couplings in which each term has at most two derivatives, \eg $R^2$. In this case, one may not have the strong constraint $\cL^{wT}-\cL^t=0$, however, using   integration by part on the left-hand side of \reef{LL}, one can fix the unknown coefficients to satisfy the constraint.   In this paper we use this latter possibility.

\section{Couplings without B-field }

In this section we are going to apply the above T-duality constraint to find the world volume couplings of dilaton and graviton at order $\alpha'^2$. Since the off-diagonal components of metric transforms to B-field under the T-duality transformations,   we assume then the metric is diagonal and B-field is zero. Using  the mathematica package ``xAct'' \cite{CS}, one can easily write all couplings with structures $R^2$, $R(\nabla\phi)^2$, $R\nabla\nabla\phi$, $(\nabla\nabla\phi)^2$, $\nabla\nabla\phi(\nabla\phi)^3$ and $(\nabla\phi)^4$ where $R$ stands for  scalar, Ricci and Reimann curvatures. In general, one should consider also the couplings involving the second fundamental form. However, such couplings are zero for O-plane. In writing the above  couplings explicitly, one should use both $G^{\mu\nu}$ and $\tG^{\mu\nu}$ for contracting the indices.  There are too many of such couplings to be able to write them here.  

To find the unknown coefficients of these couplings, we first consider the case that brane is along the $y$-direction. We then reduce it to the 9-dimensional space and  use the T-duality transformation
\beqa
\phi\rightarrow\phi-\frac{1}{2}\varphi\,\,\,,\,\,\varphi\rightarrow -\varphi
\,\,\,,  \labell{T3}
\eeqa
The resulting couplings must be the reduction  of the couplings when brane is orthogonal to the $y$-direction, \ie  \reef{LL}. This constraint  produces many equations for  the unknown coefficients. 
  Since we are interested in finding an action which is invariant under the T-duality transformations, we   use integration by part to reduce the constraints to independent ones. To do this last step, we write the curvatures and the covariant derivatives in terms of the 9-dimensional metric and then use the integration by part to find the independent structures. For example, the following terms are a total derivative:
\beqa
	&&\int d^p e^{-\phi}\sqrt{-\tg}\bigg[ \frac{1}{2}g^{\ta \tb}g^{\tc \td}g^{\te\tf}\prt_\ta\varphi\prt_\tb\varphi\prt_\tc\varphi
    \prt_\td g_{ \te\tf} -  g^{\ta \tb}g^{\tc \td}\prt_\ta\phi
   \prt_\tb\varphi\prt_\tc\varphi\prt_\td\varphi +2
 g^{\ta \tb}g^{\tc \td}\prt_\ta\varphi\prt_\tc\varphi\prt_\td\prt_\tb\varphi\nonumber\\&&
+g^{\ta \tb}g^{\tc \td}\prt_\ta\varphi\prt_\tb\varphi\prt_\td\prt_\tc\varphi -g^{\ta \tb}g^{\tc \td}g^{\te \tf}\prt_\ta\varphi\prt_\tc\varphi
   \prt_\te\varphi\prt_\tf g_{\tb \td} -g^{\ta \tb}g^{\tc\td}g^{\te \tf}\prt_\ta\varphi\prt_\tb\varphi\prt_\tc\varphi
   \prt_\tf g_{ \td \te} \bigg]\nonumber
	\eeqa	
	So one can use it to write $ g^{\ta \tb}g^{\tc \td}\prt_\ta\varphi\prt_\tc\varphi\prt_\td\prt_\tb\varphi$ in terms of other terms above. In this way the constraint imposed by the coefficient of  $ g^{\ta \tb}g^{\tc \td}\prt_\ta\varphi\prt_\tc\varphi\prt_\td\prt_\tb\varphi$ can be written in terms of constraints imposed by other  terms. One has to use such total derivative terms to reduce the constraints to independent ones. Then the coefficients of all independent structures must be zero.  
	
	Another set of constraints on the coefficients of the couplings for O-plane is that in the static gauge the curvatures and the covariant derivatives of dilaton with odd number of transverse indices must be zero.  In the 9-dimensional space, the orientifold projection is $\prt_i\varphi=\prt_i\prt_{\ta}\varphi=\prt_i\phi=\prt_i\prt_{\ta}\phi=0$ and $\prt_i\prt_{\ta}g_{\tb\tc}=\prt_i\prt_{\ta}g_{jk}=\prt_i g_{\ta\tb}=\prt_i g_{jk}=0$.
	
	Using the above constraints, all of the coefficients can be written in terms of a few constants. They produce  three type of terms which are invariant under the T-duality transformation \reef{T3}. One type is the couplings which are zero using the cyclic symmetry of the Riemann curvature, \ie
	\beqa
\cL  \supset C_1\bigg[R_{acbd}R^{abcd}- \frac{1}{2} R_{abcd}R^{abcd}\bigg]+C_2\bigg[R_{ajbi}R^{aibj}+\frac{1}{2} R_{abij} R^{abij}-R_{aibj} R^{aibj}\bigg]+\cdots
	\eeqa
where we have written the spacetime indices in terms of world-volume and transverse indices.	Since the above terms are identities, it is safe to set their coefficients to zero. One may use the above  identities  to simplify the final result. 

Another type of terms is the couplings which are total derivatives, \ie
	\beqa
\cL  &\!\!\!\supset	\!\!\!&C_3\bigg[\nabla_a\phi \nabla^a\phi \nabla_b\nabla^b\phi-\nabla_a\phi \nabla^a\phi \nabla_b\phi \nabla^b\phi+2 \nabla^a\phi \nabla_b\nabla_a\phi \nabla^b\phi\bigg]+C_4\bigg[\nabla_a\nabla^a\phi \nabla_b\nabla^b\phi\nonumber\\
	&&-\nabla_a\phi \nabla^a\phi \nabla_b\nabla^b\phi+\nabla^a\phi \nabla_b\nabla_a\phi \nabla^b\phi-\nabla_b\nabla_a\phi\nabla^b\nabla^a\phi-R^{ab}{}_{a}{}^c \nabla_b\phi \nabla_c\phi\bigg]+\cdots
	\eeqa
In the action, they can be ignored, so  it is safe to set these constants to zero too. One may  also use these total derivative terms to simplify the final result.

	The remaining T-duality invariant multiplets are the following:
	\beqa
\cL  &\!\!\!\supset	\!\!\!&	C_5\bigg[R_{ab} R^{ab}+R_{ij}R^{ij}+4 R^{ab} \nabla_b\nabla_a\Phi +4 \nabla_a\nabla^a\Phi  \nabla_b\nabla^b\Phi -4 \nabla_a\Phi  \nabla^a\Phi  \nabla_b\nabla^b\Phi\nonumber\\
	&&+4 \nabla^a\Phi  \nabla_b\nabla_a\Phi  \nabla^b\Phi -4 R^{ab}{}_a{}^c \nabla_b\Phi  \nabla_c\Phi +4 R^{ij} \nabla_j\nabla_i\Phi +4 \nabla_j\nabla_i\Phi  \nabla^j\nabla^i\Phi\bigg]\nonumber\\&&
	+C_6\bigg[R^a{}_a R^b{}_b+6 R^a{}_a \nabla_b\nabla^b\Phi +9 \nabla_a\nabla^a\Phi  \nabla_b\nabla^b\Phi -8 \nabla_a\Phi  \nabla^a\Phi  \nabla_b\nabla^b\Phi -4 R^a{}_a \nabla_b\Phi  \nabla^b\Phi \nonumber\\&&+8 \nabla^a\Phi  \nabla_b\nabla_a\Phi  \nabla^b\Phi +2 R^a{}_a \nabla_i\nabla^i\Phi +6 \nabla_a\nabla^a\Phi  \nabla_i\nabla^i\Phi -4 \nabla_a\Phi  \nabla^a\Phi  \nabla_i\nabla^i\Phi \nonumber\\&&+\nabla_i\nabla^i\Phi  \nabla_j\nabla^j\Phi \bigg]+C_7\bigg[R^a{}_a  R+3  R \nabla_a\nabla^a\phi-2  R \nabla_a\phi \nabla^a\phi+4 R^a{}_a \nabla_b\nabla^b\phi\nonumber\\&&+12 \nabla_a\nabla^a\phi \nabla_b\nabla^b\phi-12 \nabla_a\phi \nabla^a\phi \nabla_b\nabla^b\phi-4 R^a{}_a \nabla_b\phi \nabla^b\phi+16 \nabla^a\phi \nabla_b\nabla_a\phi \nabla^b\phi\nonumber\\&&+4 R^a{}_a \nabla_i\nabla^i\phi+ R \nabla_i\nabla^i\phi+16 \nabla_a\nabla^a\phi \nabla_i\nabla^i\phi-12 \nabla_a\phi \nabla^a\phi \nabla_i\nabla^i\phi+4 \nabla_i\nabla^i\phi \nabla_j\nabla^j\phi\bigg]\nonumber\\&&
	C_8\bigg[R^2+8 R\nabla_a\nabla^a\phi-8 R\nabla_a\phi\nabla^a\phi+16\nabla_a\nabla^a\phi\nabla_b\nabla^b\phi-16\nabla_a\phi\nabla^a\phi\nabla_b\nabla^b\phi\nonumber\\&&+32\nabla^a\phi\nabla_b\nabla_a\phi\nabla^b\phi+8 R\nabla_i\nabla^i\phi+32\nabla_a\nabla^a\phi\nabla_i\nabla^i\phi-32\nabla_a\phi\nabla^a\phi\nabla_i\nabla^i\phi\nonumber\\&&+16\nabla_i\nabla^i\phi\nabla_j\nabla^j\phi\bigg]+C_9R_{abij} R^{abij}+C_{10}\bigg[R_{abcd} R^{abcd}-2 R^{ab}{}_a{}^c R_{bd}{}_{cd}+2 R^{ai}{}_a{}^j R^b{}_{ibj}\nonumber\\&&-2 \nabla_b\nabla_a\phi \nabla^b\nabla^a\phi-4 R^{ab}{}_a{}^c\nabla_c\nabla_b\phi+4 R^{ai}{}_a{}^j \nabla_j\nabla_i\phi+2 \nabla_j\nabla_i\phi \nabla^j\nabla^i\phi\bigg]\nonumber
	\eeqa
The above multiplets are invariant under the simplified T-duality transformations \reef{T3}. The coefficients $C_5,\cdots, C_{10}$ should satisfy further constraint if one includes the B-field and demands that the couplings to be invariant under the full T-duality transformations \reef{T2}. 	The dilaton couplings in the multiplet with coefficient $C_{10}$ is exactly the couplings in \reef{DBI2} which are reproduced by the corresponding S-matrix element \cite{Garousi:2009dj}. A a result, we expect the invariance under the full T-duality   \reef{T2} constraints the coefficients  of other multiplets which include the dilaton, to be zero, \ie 
\beqa
C_5=C_6=C_7=C_8=0
\eeqa
 In the next section, we use the above constraint and include the B-field couplings to the multiplets $C_9$ and $C_{10}$.

\section{Couplings with B-field } 

In this section we are going to find the connection between the constants $C_9,\, C_{10}$ and find the unknown coefficients of  B-field couplings by constraining  the whole couplings to be invariant under the T-duality transformation \reef{T2}. The multiplets with coefficient $C_9$ and $C_{10}$ in terms of 10-dimensional   indices are the following:
\beqa
\cL&\!\!\!\!\!\!\!\supset	\!\!\!\!\!\!\!&	C_9\bigg[2R_{\alpha \beta \gamma \delta }R^{\alpha \beta \gamma \delta }-4R_{\alpha \gamma \beta \delta }R^{\alpha \beta \gamma \delta }-4R_\alpha {}^{\gamma \delta \epsilon }R_{\beta \gamma \delta \epsilon } \tG^{\alpha \beta }+8R_\alpha{}^{ \gamma \delta \epsilon }R_{\beta \delta \gamma \epsilon } \tG^{\alpha \beta }\nonumber\\&&
 +2R_\alpha{}^ \epsilon{}_ \gamma{}^ \varepsilon R_{\beta \epsilon \delta \varepsilon }\tG^{\alpha \beta } \tG^{\gamma \delta }-2R_\alpha {}^\epsilon {}_\gamma {}^\varepsilon R_{\beta \varepsilon \delta \epsilon } \tG^{\alpha \beta } \tG^{\gamma \delta }-4R_{\alpha \gamma \epsilon }{}^\zeta R_{\beta \varepsilon \delta \zeta } \tG^{\alpha \beta } \tG^{\gamma \delta } \tG^{\epsilon \varepsilon}\nonumber\\&&
+2R_{\alpha \gamma \epsilon \zeta }R_{\beta \varepsilon \delta \eta } \tG^{\alpha \beta } \tG^{\gamma \delta } \tG^{\epsilon \varepsilon } \tG^{\zeta \eta}\bigg]+ C_{10}\bigg[9R_{\alpha \beta \gamma \delta }R^{\alpha \beta \gamma \delta }-18R_{\alpha \gamma \beta \delta }R^{\alpha \beta \gamma \delta }-16R_\alpha {}^{\gamma \delta \epsilon }R_{\beta \gamma \delta \epsilon } \tG^{\alpha \beta }\nonumber\\&&
+32R_\alpha {}^{\gamma \delta \epsilon }R_{\beta \delta \gamma \epsilon } \tG^{\alpha \beta }-8R_{\alpha \gamma} {}^{\epsilon \varepsilon }R_{\beta \epsilon \delta \varepsilon } \tG^{\alpha \beta } \tG^{\gamma \delta }+8R_\alpha {}^\epsilon {}_\gamma {}^\varepsilon R_{\beta \epsilon \delta \varepsilon} \tG^{\alpha \beta } \tG^{\gamma \delta }-8R_\alpha {}^\epsilon{}_\gamma {}^\varepsilon R_{\beta \varepsilon \delta \epsilon }\tG^{\alpha \beta } \tG^{\gamma \delta}\nonumber\\&&
+2R_\alpha {}^\epsilon {}_\beta {}^\varepsilon R_{\gamma \epsilon \delta \varepsilon } \tG^{\alpha \beta } \tG^{\gamma \delta }+4R_{\alpha \gamma \beta}{}^ \zeta R_{\delta \epsilon \varepsilon \zeta }\tG^{\alpha \beta } \tG^{\gamma \delta } \tG^{\epsilon \varepsilon }+2R_{\alpha \gamma \epsilon \zeta }R_{\beta \varepsilon \delta \eta } \tG^{\alpha \beta } \tG^{\gamma \delta } \tG^{\epsilon \varepsilon } \tG^{\zeta \eta }\nonumber\\&&
+2 \nabla_\beta  \nabla_\alpha \phi \nabla^\beta \nabla^\alpha \phi-4 \tG^{\alpha \beta } \nabla_\gamma  \nabla_\beta  \phi  \nabla^\gamma  \nabla_\alpha \phi\nonumber\\&&+4R_\alpha {}^\gamma{}_\beta {}^\delta  \tG^{\alpha \beta } \nabla_\delta  \nabla_\gamma \phi-8R_{\alpha \gamma \beta}{}^ \epsilon  \tG^{\alpha \beta} \tG^{\gamma \delta} \nabla_\epsilon \nabla_\delta \phi\bigg]\nonumber
\eeqa
We have to add to the above couplings, the B-field couplings in which the tensors with structures $H^4$, $RH^2$, $\nabla\phi\nabla\phi H^2$ and $\nabla\nabla\phi H^2$ contract with $G^{\mu\nu}$ or $\tG^{\mu\nu}$. Again there are too many of such couplings to be able to write them here. To find their coefficients we impose the T-duality constraint \reef{LL}. 

To simplify the calculation, we assume that the base fields $g_{\alpha\beta}=\eta_{\alpha\beta}$ and $b_{\alpha\beta}=0$. With this assumption, there are still too many terms to handle even with computer. Since the overall factor of all terms is $\int d^px\,e^{-\phi}\sqrt{-\det(g_{\ta\ta})}$ which is independent of $g_\ta,\, b_\ta$, we can consider the cases that the number of  $g_\ta,\, b_\ta$ to be 0, 2, 4 and  6 separately. In each case one can use integration by part to find independent structures whose coefficients must be zero. In the case that the number of  $g_\ta,\, b_\ta$ is 0, one finds no constraint because it is a simplified version of the previous section in which $g_{\alpha\beta}=\eta_{\alpha\beta}$. In the case that the number of $g_\ta,\, b_\ta$ is odd or is larger than 6, there is no constraint, \ie the left hand side of \reef{LL} is zero.   Moreover, for the case that the number of $g_\ta,\, b_\ta$ is 6, there is no term which has derivative of dilaton or $\varphi$, \ie all four derivatives are on $g_{\ta}$ or $b_{\ta}$. In this case, the strong  condition $\cL^{wT}-\cL^t=0$ produces  only independent constraints because the total derivative terms contains, among other things, the  derivatives of dilaton which are not in the list of constraints produced by $\cL^{wT}-\cL^t=0$.  

 For O-plane action, the   covariant derivatives of B-field  with even number of transverse indices are projected out. So the coefficient of such term constraint to be zero. In the 9-dimensional space, the orientifold projection is $\prt_i\varphi=\prt_i\prt_{\ta}\varphi=\prt_i\phi=\prt_i\prt_{\ta}\phi=0$, $\prt_{\ta}b_i=\prt_ib_{\ta}=\prt_i\prt_jb_k=\prt_i\prt_{\ta}b_{\tb}=0$ and  $\prt_{\ta}g_\tb=\prt_ig_j=\prt_i\prt_\ta g_j=\prt_i\prt_jg_{\ta}=0$.

Using the above constraints, all of the coefficients can be written in terms of a few constants. In particular, the T-duality   fixes $C_9=-C_{10}$ which makes the above Lagrangian to be proportional to the Lagrangian in \reef{DBI2}. The non-zero coefficients of the B-field produce  two type of terms which are invariant under the T-duality transformation \reef{T2}. One type is the couplings which are zero using the cyclic symmetry of the Riemann curvature or the Bianchi identity of B-field, \ie
\beqa
\cL  &\!\!\!\!\!\supset	\!\!\!\!\!&a_1\bigg[-12 H_{i}{}^{lm} H^{ijk} H_{jl}{}^n H_{kmn}+9 H_{ij}{}^l H^{ijk} H_{k}{}^{mn} H_{lmn}-H_{ijk} H^{ijk} H_{lmn} H^{lmn}\bigg]\nonumber\\&&
+a_2\bigg[-\nabla_ b H_{aci} \nabla^cH^{abi}+\frac{1}{2} \nabla_c H_{ a b i} \nabla^c H^{ a b i}-\frac{1}{6} \nabla_i H_{ a b c} \nabla^i H^{abc}\bigg]\labell{iden1}\\&&
+a_3\bigg[\frac{1}{2} H^{abi} H^{cd}{}_i R_{abcd}-H^{abi} H^{cd}{}_i R_{acbd}\bigg]+a_4\bigg[\frac{1}{3} \nabla_a H_{ i j k} \nabla^a H^{ijk}-\nabla^a H^{ijk} \nabla_k H_{ a i j}\bigg]+\cdots\nonumber
\eeqa
These coefficients can be set to zero.  The remaining terms have only one overall unknown coefficient, \ie
\beqa
\cL  &\!\!\!\supset	\!\!\!&C_{10}\bigg[\frac{1}{8} H_a{}^{cj} H^{abi} H_b{}^d{}_j H_{cdi}+\frac{1}{4} H_{ a b}{}^jH^{abi} H_{ c d j} H^{cd}{}_i-\frac{1}{6} H_a{}^{cj} H^{abi} H_{ b c}{}^k H_{ i j k}\nonumber\\&&-\frac{1}{4} H_{ab}{}^j H^{abi} H_i{}^{kl} H_{ j k l}
+\frac{1}{32} H_{ i j}{}^l H^{ijk} H_ k{}^{mn} H_{ l m n}-\frac{1}{288} H_{ i j k} H^{ijk} H_{ l m n} H^{lmn}\nonumber\\&&-\frac{3}{4} H^{abi} H^{cd}{}_ i R_{ a b c d}+\frac{1}{2} H^{abi} H_ i{}^{jk} R_{ a b j k}
-\frac{3}{2} H^{abi} H^{cd}{}_ i R_{ a c b d}+H^{abi} H_i{}^{jk} R_{ a j b k}+R_{ a b c d} R^{abcd}\nonumber\\&&-R_{ a b i j} R^{abij}+\frac{1}{2} H_{ i j}{}^l H^{ ijk}R^a{}_{kal}
-2 H_ a{}^{cj} H^{abi} R_{ b c i j}+2 H_ a{}^c{}_ i H^{abi} R_{ b}{}^d{}_{ c d}-2 R^{ab}{}_{ a}{}^c R_ b{}^d{}_{ c d}\nonumber\\&&+H_ a{}^{cj} H^{abi} R_{ b i c j}-H_ a{}^{cj} H^{abi} R_{ b j c i}
+2 R^{ai}{}_a{}^j R^b{}_{ i b j}-\frac{3}{2} H_{ a b}{}^j H^{abi} R^c{}_{ i c j}+\frac{1}{4} H_ i{}^{lm} H^{ijk} R_{ j k l m}\nonumber\\&&-\frac{1}{2} H_ i{}^{lm} H^{ijk} R_{ j l k m}-\frac{1}{6} \nabla_ a H_{ i j k} \nabla^a H^{ijk}
-2 \nabla_ b \nabla_ a \phi \nabla^b\nabla^a\phi+\nabla_aH^{abi} \nabla_cH_b{}^c{}_i\nonumber\\&&-H^{abi}\nabla_a\phi \nabla_cH_b{}^c{}_i+H_a{}^c{}_i H^{abi} \nabla_c\nabla_b\phi-4 R^{ab}{}_a{}^c \nabla_c\nabla_b\phi
+3 \nabla_bH_{ a c i} \nabla^cH^{abi}\nonumber\\&&-\frac{1}{2} \nabla_cH_{ a b i} \nabla^cH^{abi}-H^{abi} \nabla_bH_{ a c i} \nabla^c\phi
-\frac{3}{2} H_{ a b}{}^j H^{abi} \nabla_j\nabla_i\phi
+4 R^{ai}{}_{a}{}^j \nabla_ j \nabla_ i\phi\nonumber\\&&+2 \nabla_j\nabla_i\phi \nabla^j\nabla^i\phi+\frac{1}{2} H_{ i j}{}^l H^{ijk}\nabla_l\nabla_k\phi\bigg]\labell{final}
\eeqa
The overall constant $C_{10}$ can be fixed by comparing $R^2$ terms above with the corresponding terms in the action \reef{DBI}. The above result can be further simplified using the Bianchi identities in \reef{iden1}.   

Since we have used the assumption  $g_{\alpha\beta}=\eta_{\alpha\beta}$, the calculation in this section  could not find the couplings that are total derivatives. However, using the identity 
\beqa
H^\alpha{}_{  d i} R_{ b a c \alpha }-H^\alpha{}_{  c i} R_{ b a d \alpha }+H^\alpha{}_{ c d} R_{ b a i \alpha }+\nabla_a\nabla_bH_{ c d i}-\nabla_b\nabla_aH_{ c d i}=0
\eeqa
One can verify that the following is a  total derivative term:
\beqa
&&\int d^{p+1}x e^{-\phi}\sqrt{-\tG}\bigg[-\nabla_aH^{abi} \nabla_cH_b{}^c{}_i+ H_a{}^{cj} H^{abi} R_{ b c i j}+H^{abi} H^e{}_{ c i} R_b{}^c{}_{ae}\nonumber\\&&-H_a{}^c{}_i H^{abi}R_b{}^d{}_{ c d}-H^{abi} \nabla_b\phi \nabla_cH_a{}^c{}_i
-\nabla_bH_{ a c i} \nabla^cH^{abi}+H^{abi}\nabla_ b H_{ a c i} \nabla^c\phi\bigg]
\eeqa
 Using the above total derivative term and the Bianchi identities in \reef{iden1}, we have found that  the T-dual couplings in \reef{final} can be simplified to the couplings in \reef{DBI2}, \reef{LDBI} and \reef{finalH}. 

Our calculation indicates  that the derivatives of dilaton  in the world volume theory, \ie equations \reef{DBI2}, \reef{LDBI} and \reef{finalH},  appear only through the replacement $\hR_{\mu\nu}\rightarrow \cR_{\mu\nu}$. Such replacement appears also in the Chern-Simons part   at the quadratic order fields \cite{Garousi:2011fc}. In fact $\cR_{\mu\nu}$ is invariant under linear T-duality. We expect, apart from the overall dilaton factor $e^{-\phi}$, dilaton appears in the world volume theory only through this replacement. 

We have seen that the couplings in \reef{DBI2}, \reef{LDBI} and \reef{finalH} are invariant under the full T-duality transformation \reef{T2}. However, these couplings are not invariant under  S-duality for O$_3$-plane. The S-duality   requires adding appropriate RR couplings. We expect all RR couplings can be found by requiring the world volume action to be invariant under both T-duality and S-duality. The consistency of the couplings under S-duality and linear T-duality has been considered in  \cite{Garousi:2011fc} to find  quadratic world-volume couplings. It would be interesting to extend the calculation in \cite{Garousi:2011fc} to full nonlinear T-duality to find all NSNS and RR couplings on the world volume of O-plane. It would be also interesting to confirm the cubic couplings in \reef{finalH} by the corresponding S-matrix element of three closed string states at projective plane level.  Two-point function on projective plane has been calculated in \cite{Garousi:2006zh}.

{\large Note added}: During the completion of this work,   the preprint \cite{Robbins:2014ara} appeared which has some overlaps with the results in this paper.

{\bf Acknowledgments}:   This work is supported by Ferdowsi University of Mashhad under grant 2/32987(1393/11/21). 
\bibliographystyle{/Users/Nick/utphys} 
\bibliographystyle{utphys} \bibliography{hyperrefs-final}


\providecommand{\href}[2]{#2}\begingroup\raggedright

\endgroup

\end{document}